# Spatiotemporal Assessment of Aircraft Noise Exposure Using Mobile Phone-Derived Population Estimates and High-Resolution Noise Measurements

Soohwan Oh[a], Hyunsoo Cho[b], Jungwoo Cho[c*]

[a] *Department of Aerospace Industrial & Systems Engineering, Hanseo University, 46. Hanseo 1-ro, Chungcheognam-do, 31962, Republic of Korea*

[b] *Korea Airports Corporation, 78 Haneul-gil Gangseo-gu, Seoul, 07505, Republic of Korea*

[c] *Department of Air and Space Transportation, Korea Transport Institute, 370 Sicheon-daero, Sejong, 30147, Republic of Korea*

[*] *Corresponding Author*

*Email addresses: suhwan@hanseo.ac.kr (S. Oh), jungwoo@koti.re.kr (J. Cho).*

# Spatiotemporal Assessment of Aircraft Noise Exposure Using Mobile Phone-Derived Population Estimates and High-Resolution Noise Measurements

**ABSTRACT**

Aircraft noise exposure has traditionally been assessed using static residential population data and long-term average noise metrics, often overlooking the dynamic nature of human mobility and temporal variations in operational conditions. This study proposes a data-driven framework that integrates high-resolution noise measurements from airport monitoring terminals with mobile phone-derived de facto population estimates to evaluate noise exposure with fine spatio-temporal resolution. We develop hourly noise exposure profiles and quantify the number of individuals affected across regions and time windows, using both absolute counts and inequality metrics such as Gini coefficients. This enables a nuanced examination of not only who is exposed, but when and where the burden is concentrated. At our case study airport, operational runway patterns resulted in recurring spatial shifts in noise exposure. By incorporating de facto population data, we demonstrate that identical noise operations can yield unequal impacts depending on the time and location of population presence, highlighting the importance of accounting for population dynamics in exposure assessment. Our approach offers a scalable basis for designing population-sensitive noise abatement strategies, contributing to more equitable and transparent aviation noise management.

# 1 INTRODUCTION

The aviation industry is one of the fastest-growing sectors globally, generating substantial economic benefits while also posing a range of environmental challenges (Tveter, 2017; Jin et al., 2020; Tirtha et al., 2023). This growth trajectory is expected to continue in tandem with the expansion of the global economy (Franz et al., 2022; Tirtha et al., 2022). However, the rapid increase in air traffic has intensified concerns regarding noise pollution, especially in densely urbanized areas where airports are in close proximity to residential communities (Grampella et al., 2016; Postorino and Mantecchini, 2016; Gagliardi et al., 2018). Evidence links the chronic disturbance caused by aircraft noise to adverse impacts on residents' well-being, ranging from mild annoyance and sleep disruption to the exacerbation of community opposition toward nearby aviation facilities (Babisch et al., 2009; Perron et al., 2012; Lefevre et al., 2020).

In an effort to mitigate these concerns, the International Civil Aviation Organization (ICAO) introduced the Balanced Approach, a framework that underscores strategies including noise reduction at the source, land-use planning, noise abatement operational procedures, and operating restrictions (ICAO, 2008). Although many airports have adopted measures such as curfews, preferential runway usage, and noise limits under this framework (Girvin, 2009; Porter, 2017, 2023), the practical effectiveness of these efforts depends on the accuracy of noise exposure assessments, which not only inform regulatory and operational decisions but also shape community engagement strategies.

A key element in any aircraft noise management plan is the estimation of how many people experience potentially harmful noise exposure (Ganic et al., 2023). Traditionally, this has been done by overlaying annual noise contours on static census data to estimate exposure – implicitly assuming that individuals spend most of their time at home (Ganic et al., 2023). In reality, daily mobility patterns often deviate from this assumption, especially in large urban regions where residents may travel frequently for work, leisure, or airport-related activities. Consequently, relying on census data may under- or overestimate the populations actually exposed to harmful noise levels.

Further complicating this issue is the gap that can arise between modeled and measured noise. Even with sophisticated software tools that account for aircraft type, engine characteristics, flight paths, and meteorological conditions, models may fail to capture local variations in atmospheric effects or flight patterns (Simons et al., 2022).

To address these limitations, the present study introduces a data-driven approach for assessing aircraft noise exposure in a dynamic urban context. Our approach incorporates high-resolution population estimates derived from mobile phone data to represent the spatiotemporal variability in how people circulate through the city. In parallel, we incorporate finely resolved noise data – collected at short intervals from multiple noise monitoring terminals – to capture local acoustic variations more accurately. We then employ an ensemble machine learning model and SHAP (Shapley Additive Explanations) to interpret the model outputs and align them with established domain knowledge. We integrate these components to generate an hourly, census tract level exposure indicator. Beyond the total number of people affected, we also quantify inequality in exposure distributions using the Gini coefficient.

The remainder of this paper is structured as follows: Section 2 presents the literature review; Section 3 outlines the data and methods utilized in our study. In Section 4, we present and discuss the results obtained. Finally, Section 5 offers concluding remarks on our study's findings and implications.

## 2 LITERATURE REVIEW

Accurately assessing aircraft noise exposure requires robust modeling of noise emissions and an understanding of the dynamic human context in which those emissions are experienced. Previous research has primarily focused on refining noise estimation models, ranging from physical acoustic simulations to data-driven machine learning approaches, but often fails to capture the temporal and spatial variability of human exposure. In light of these gaps, this section outlines three key domains relevant to our study: (1) noise estimation methods and limitations, (2) integration of human mobility in aircraft noise assessments, and (3) regulatory and public health approaches to noise exposure.

## 2.1. Noise estimation models and limitations

Over the past two decades, aircraft noise models have evolved significantly. Physical models, grounded in acoustic propagation theories, estimate noise levels based on detailed aircraft characteristics, air traffic operational procedures, and environmental factors (Filippone, 2014; Guo et al., 2019). While such models offer high theoretical accuracy, they are often constrained by proprietary input parameter, scarcity of operational data and limited adaptability to real-time airport environments.

Conversely, best-practice models offer a simplified approach by treating aircraft as single-point sources, allowing for quick assessments using basic operational data (Ang and Cui, 2022). These models are influential in regulatory contexts, such as land-use planning, but often underestimate daily noise variation, particularly in areas with variable air traffic or weather (Simons et al., 2022). They also do not fully account for local noise variations, leading to spatial inaccuracies, particularly under real-time or non-standard flight conditions.

Recent advancements in data-driven machine learning models offer new potential to refine noise estimations. Linear regression and neural network models trained on flight and noise monitoring data have shown promising results in predicting local noise levels under diverse conditions (Gagliardi et al., 2018; Zellmann et al., 2018). Notably, neural networks have improved the modeling of nonlinear interactions, such as the combined effects of aircraft type, trajectory, and weather (Revoredo et al., 2016; Tenney et al., 2020; Vela and Oleyaei-Motlagh, 2020). Despite their empirical success, many machine learning models remain black boxes, lacking interpretability and offering limited guidance for actionable airport policy or community engagement.

## 2.2. Integration of human mobility in aircraft noise assessments

Traditional exposure assessments typically overlay static census data with annual noise contours, assuming people remain in one place throughout the day (Nguyen et al., 2023). This method assumes that residents spend the majority of their time at home, but this assumption is increasingly inaccurate in modern urban environments where daily mobility patterns are highly dynamic (Ganic et al., 2023). As a result, traditional assessments often misestimate exposure, particularly in areas with significant commuter traffic or frequent changes in population density.

Recent efforts have begun to address this limitation. Ganic et al. (2023) introduced a dynamic noise mapping framework using activity-based travel survey data from the London Travel Demand Survey (LTDS) to capture intra-day population mobility and assess its impact on noise exposure. While their work represents a meaningful advancement beyond static census-based methods, they emphasized that future research should incorporate higher-resolution human mobility datasets, such as those derived from mobile phone signals, to further improve exposure realism.

Despite the growing recognition of mobility's role in exposure, most exposure estimation models remain static, failing to integrate high-resolution real-time population data. Our study addresses this gap by leveraging mobile phone-derived population estimates to capture the temporal fluctuations of human presence across different geographic areas and integrating this data with advanced noise estimation models.

## 2.3. Regulatory and public health approaches to noise exposure

A growing body of international guidance underscores the importance of quantifying noise exposure not only in terms of acoustic metrics, but also through the lens of public health and population equity. This evolution is reflected in the regulatory frameworks and guidelines issued by ICAO, and European Union (EU), and World Health Organization (WHO).

The ICAO Balanced Approach to Aircraft Noise Management (ICAO, 2008) provides the foundational strategy for global airport noise policy. It is structured around four pillars: reduction of noise at the source, land-use planning and management, noise abatement operational procedures, and operating restrictions. Central to ICAO's exposure assessment methodology is the use of annual average noise contours overlaid with static residential population data. While ICAO recognizes that population distribution around airports varies throughout the day, it does not incorporate dynamic population data into standard noise exposure calculations.

The European Union's Environmental Noise Directive (Directive 2002/49/EC) formalized requirements for member states to assess and manage environmental noise. The directive mandates the use of noise indicators such as $L_{den}$ and $L_{night}$ to estimate long-term exposure (European Union, 2002). However, it initially left the health-risk quantification of noise exposure relatively implicit. In 2020, the European Commission adopted Directive (EU) 2020/367, an annex to the 2002 directive, which introduced does-response functions to quantify the risk of harmful effects from noise exposure (European Commission, 2020). Specifically, the annex outlines statistical models to estimate the percentage of individuals highly annoyed or suffering from sleep disturbances as a function of noise metrics such as $L_{den}$ and $L_{night}$. These functions enable policymakers to move beyond acoustic levels and directly estimate the population-level health risks of noise exposure.

Parallel to these developments, the WHO set evidence-based health thresholds for various noise sources, including aircraft (WHO, 2018). The WHO recommends that $L_{den}$ levels from aircraft should not exceed 45 dBA to avoid high annoyance and $L_{night}$ should remain below 40 dBA to protect against sleep disturbance. These thresholds were derived from epidemiological studies linking long-term noise exposure to cardiovascular disease, cognitive impairment, and reduced well-being.

In addition, operational trials at London Heathrow Airport have demonstrated that community perceptions of aircraft noise are shaped not only by sound levels but also by the timing, duration, and predictability of quiet periods, often referred to as "respite" (Porter, 2017, 2023). Notably, research indicates that a 4–9 dB reduction is typically required for noticeable detection, while a greater than 9 dB reduction is widely perceived by residents as a valuable break from noise (Porter, 2023). Even modest reductions smaller than 4 dB have been associated with improved community attitudes under real-world conditions. These studies also highlight that residents place the highest subjective value on early morning (07:00–11:00) and evening (19:00–23:00) quiet periods, emphasizing the importance of temporal regularity and predictability. These insights align with broader evidence suggesting that non-acoustic factors, including transparency, procedural fairness, and community trust, can be as influential as sound metrics in determining public acceptance of noise mitigation policies. Together, these findings underscore the policy relevance of characterizing aircraft noise exposure using hourly, population-weighted indicators—a gap that this study seeks to address.

In summary, the abovementioned international frameworks mark a shift toward more health-oriented and socially responsive noise policy. Yet most existing implementations continue to rely on static population datasets and annualized noise contours, which obscure the dynamic realities of exposure in urban settings. Our study contributes to this shift by developing a high-resolution, temporally dynamic noise exposure model that integrates mobile-phone-derived population estimates and fine-grained airport noise measurements. In doing so, we generate a new class of exposure metrics that reflect not only the magnitude of noise but also its distribution and impact by location and hour of day.

## 3 METHODOLOGY

This study proposes a novel, data-driven indicator to assess aircraft noise exposure in a highly dynamic urban environment, integrating both population mobility data and high-resolution noise measurements. In this section, we describe the datasets used, the noise estimation approach, and the machine learning model employed to validate the results.

### 3.1. Data sources and preprocessing

### 3.1.1. Hourly de facto population estimates

The primary data used in this study is de facto population data, which represents the number of individuals physically present within each census tract during one-hour intervals. This dataset is derived from Korea Telecom's (KT) geolocation records, which are based on anonymized mobile phone signals collected through KT-operated base stations across Seoul. An individual's location is inferred by the most recent base station that his or her device has connected to (Ministry of Science and ICT, 2023). KT's network provides comprehensive coverage throughout Seoul, encompassing 19,153 census tracts, the city's smallest administrative unit, each with a median area of 11,689 m². The de facto population dataset is publicly available through the Seoul Big Data Campus platform (Seoul Metropolitan Government, 2025).

KT's proprietary model integrates mobile phone signals with supplementary demographic and transportation datasets to estimate the population present in each census tract. The model accounts for factors such as residential population, employed population, and public transit ridership, and adjusts population estimates based on KT's mobile market share, LTE/5G subscription rate, and the proportion of active devices. Importantly, KT's signal data are automatically recorded as devices move between base stations—regardless of whether users are actively using their phones—allowing for continuous and precise tracking of population distribution and movement (Seoul Metropolitan Government, 2025).

To address underrepresentation among demographic groups with lower mobile phone usage, such as the very young and the elderly, the model incorporates correction factors based on resident registration data (Seoul Metropolitan Government, 2025).

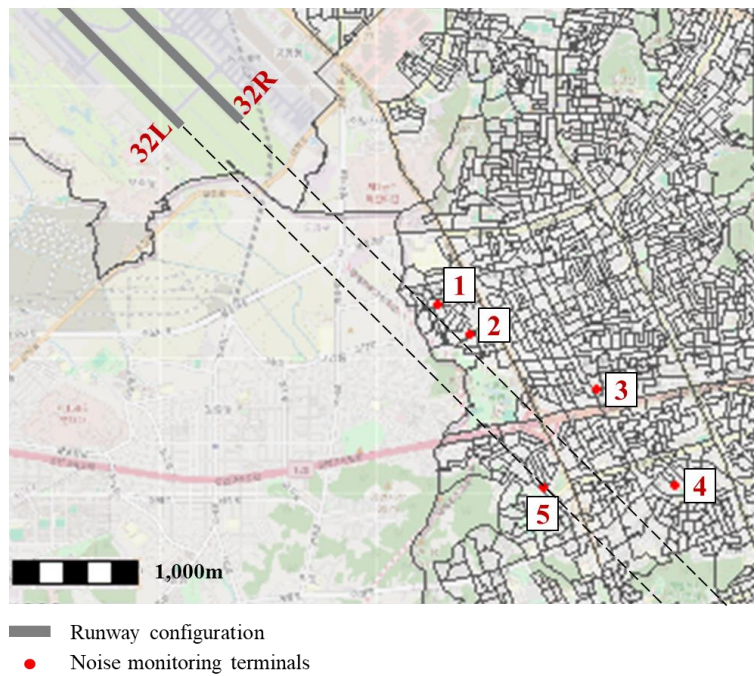

**Figure 1.** Geographical layout of NMT sites and census tract boundaries near Runways 32L and 32R of Gimpo International Airport

### 3.1.2. Airport noise, flight, and weather data

Noise data for this study were collected in January 2023 from five noise monitoring terminals (NMTs) located near Gimpo International Airport (GMP) in South Korea (see Figure 1), which is a site known for its high levels of aircraft noise. In 2019, prior to the COVID-19 pandemic, GMP handled over 25 million passengers and more than 140,000 flights, underscoring its significance as a high-traffic airport. The noise data were provided by the Korea Airports Corporation (KAC), which operates the NMTs and manages the measurement infrastructure.

Each NMT recorded sound pressure levels at three-second intervals, allowing for the detection of high-frequency noise variations that are often masked in long-term average metrics. The devices used—Brüel & Kjær Noise Monitoring Terminal models 3639-G and 3639-C—are certified to IEC 61672 Class 1 specifications and offer a dynamic range of 110 dBA. Terminal locations were strategically selected to capture diverse runway operations, including takeoff and landing directions, while minimizing interference from background noise and surrounding topography.

For analysis, the three-second noise records were aggregated into hourly equivalent continuous sound levels ($LAeq_{1h}$), a standard metric used in environmental acoustics. Only observations exceeding 60 dBA were retained, as this threshold is known to interfere with normal speech and has been associated with adverse health outcomes (Eagan, 2007).

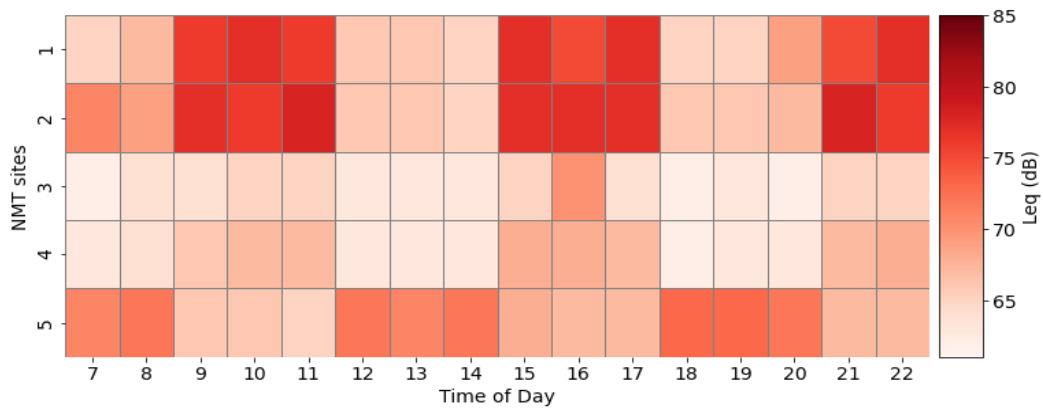

**Figure 2.** Hourly LAeq distribution across census tracts

Runway operations at GMP follow a fixed three-hour rotation schedule to provide periodic noise relief to surrounding communities. For example, from 09:00 to 11:59, Runway 32R is used for landings and 32L for departures; from 12:00 to 14:59, the configuration is reversed. This pattern may also be adjusted in response to weather conditions or operational demands. As illustrated in Figure 2, NMT 1 and 2, located directly under the Runway 32R approach path, consistently register pronounced LAeq peaks during 32R landing periods (e.g., 09:00-11:59, 15:00-17:59, and 21:00-22:59). In contrast, NMT 5, situated under the Runway 32L approach shows elevated LAeq values during 32L landing periods, while remaining lower during 32R operations. Additionally, NMTs 1 and 2 exhibit consistently higher LAeq levels compared to NMTs 3 and 4, which are positioned farther from the runways. This reflects the expected acoustic gradient associated with aircraft altitude, where closer proximity to the runway results in higher ground-level noise exposure due to lower altitude and engine operating conditions.

In addition to acoustic measurements, we incorporated flight schedule data from KAC and meteorological data from the Aviation Meteorological Office (AMO). The flight schedule dataset includes detailed information on aircraft type, engine type, airline, runway usage, and timestamped departure and arrival events. This granularity enabled us to assess the role of aircraft-engine pairings and operational configurations in shaping noise patterns. Meteorological data were recorded hourly and include variables such as temperature, wind speed, wind direction (relative to runway heading), and cloud cover. These factors are known to influence sound propagation and are thus critical for accurate noise modeling. In total, we used we used 22 variables to characterize the airport's operational environment, encompassing meteorological conditions, aircraft-engine combinations, and temporal scheduling features, which are detailed in Supplementary Table S1.

### 3.1.3. Integration of datasets

The population and noise, flight, and weather datasets were integrated at the hourly level. For each census tract, we matched the de facto population to the closest NMT's $LAeq_{1h}$ measurement, and linked this information to operational and meteorological conditions during the same hour. In the integration process, the $LAeq_{1h}$ value measured at a given NMT was used as a proxy for noise exposure across the entire census tract in which the terminal is located. This assumption is supported by the relatively small spatial extent of census tracts in the study area. While this simplification may not capture fine-grained acoustic variations in complex built environments, it offers a practical and reasonably accurate framework for tract-level exposure modeling under the constraints of existing sensor coverage. Overall, this integrated approach enabled a high-resolution reconstruction of real-world noise exposure patterns, capturing both the spatial distribution of populations at risk and the dynamic variability of the acoustic environment.

**3.2. Airport noise estimation**

To model the complex interactions between aircraft operations, meteorological conditions, and noise levels, we applied in ensemble machine learning approach using the XGBoost (eXtreme Gradient Boosting) algorithm. XGBoost was chosen for its ability to handle large, high-dimensional datasets and capture nonlinear relationships between variables effectively. The model was trained using 90% of the data, with the remaining 10% reserved for testing and validation. To prevent overfitting, we set the learning rate to 0.05 and optimized the number of boosting rounds between 10 and 1,000, following best practices (Chen and Guestrin, 2016).

In XGBoost, a sequential ensemble approach is performed to form sequential decision trees. Such tree-based algorithm allows us to take into account the dependency between independent variables. The first learner is fitted to the full input dataset, and the second learner is fitted to the residuals. The process repeats several times before it reaches the stopping point. Predictions from each learner are summed to from the final prediction. The mathematical equation of such process is as follows. $f_i^p = \sum_{k=1}^{l} f_k(x_i) = f_i^{(p-1)} + f_i(x_i)$, where $f_p(x_i)$ is the learner at phase $p$, $f_i^p$ and $f_i^{(p-1)}$ are the prediction at phase $p$ and $p-1$, and $x_i$ is the input variables. In order to prevent overfitting issues, XGBoost's objective function consists of a convex loss function and a regularization term, $Obj^{(p)} = \sum_{k=1}^{n} l(\bar{y}_i, y_i) + \sum_{k=1}^{p} \sigma(f_i)$, where $l$ is the loss function, $n$ is the number of observations, and $\sigma$ is the regularization term.

The model was trained on various input features, including flight schedule data (aircraft type, engine type, runway usage), meteorological variables (temperature, wind speed, cloud cover, wind direction), and spatial features (geographic location of the noise monitoring stations). The trained model generates noise estimates for different operational and meteorological scenarios, providing a more nuanced understanding of the factors driving noise exposure.

**3.3. Technical validation**

**3.3.1. De facto population data validation**

To evaluate the reliability of the de facto population estimates derived from mobile phone data, we compared their temporal dynamics with known land use characteristics in representative areas of Seoul. For instance, in the central business district of Gangnam-gu, we expect to observe higher population counts during daytime hours due to the influx of workers and visitors, followed by a decline at night. Conversely, in primarily residential districts such as Yangcheon-gu, population levels should be lower during work hours and higher in the evening and early morning. These expected temporal patterns provide a benchmark for assessing the face validity of the de facto population dataset.

**3.3.2. Airport noise data validation**

To assess the validity of the airport noise estimates generated by our machine learning model, we focused on the interpretability of the estimates in relation to known physical and operational principles. Specifically, we examined whether the model's most influential variables were consistent with domain knowledge of noise generation – such as the effects of meteorological conditions and aircraft-engine configurations.

Given the inherent complexity of ensemble learning algorithms like XGBoost, we employed the SHAP framework developed by Lundberg and Lee (2017) to improve transparency. Unlike traditional feature importance metrics, SHAP not only identifies influential variables but also reveals the direction and magnitude of their effects on model estimations. This dual insight allows us to interpret individual estimations and validate the overall logic of the model.

In our study, SHAP values were computed for all features across the dataset. The SHAP value can be calculated as: $\phi_i(f, x) = \sum_{S \subseteq M \setminus \{i\}} \frac{|S|!(|M|-|S|-1)!}{|M|!} [f(S \cup \{i\}) - f(S)]$, where $M$ is the set of all input variables; $S$ is a subset of $M$ with the $i$ feature excluded from $M$; $[f(S \cup \{i\}) - f(S)]$ is the marginal feature contribution $i$ variable. A positive SHAP value indicates that a given variable contributes to a higher noise level, while a negative value suggests noise reduction. For example, wind deviation from runway alignment and low cloud cover showed strong positive SHAP effects on take-off and landing noise – both findings that align with known physical phenomena such as sound reflection and dispersion. This interpretability check reinforces confidence in our model's ability to capture real-world relationships.

**3.4. Integrated noise-exposure indicator**

As mentioned earlier, traditional aircraft noise exposure assessments have primarily relied on static census population data and long-term average noise metrics. While this approach has proven useful for high-level planning, it fails to capture the spatial and temporal dynamics of real-world exposure– particularly in highly urbanized areas with significant intra-day population fluctuations. Recognizing this limitation, recent efforts by researchers (Ganic et al., 2021; Ganic et al., 2023) have prompted the use of more refined indicators that integrate both noise measurements and temporal population changes.

In this study, we propose an integrated, high-resolution noise-exposure indicator that reflects both the magnitude of noise and the spatio-temporal distribution of exposed population. Our approach leverages two core components: (1) aircraft noise levels measured on each 3-second and converted to an hourly basis, and (2) hourly de facto population estimates derived from mobile phone data. The fusion of these datasets supports a more nuanced assessment of noise exposure that reflects variations in airport operations and urban mobility.

### 3.4.1. Hourly exposure at census tract-level

To capture the number of people exposed to noise above the threshold θ in each hour, we measure the hourly exposure $E_{i,t}^{(\theta)}$ at hour $t \in \{1, ..., T\}$ be defined as: $E_{i,t}^{(\theta)} = n_{i,t} \cdot \delta(L_{i,t} > \theta)$, where $n_{i,t}$ is the number of people present in census tract $i$ at time $t$, from de facto population estimates, $L_{i,t}$ is the measured noise level in census tract $i$ at time $t$, $\theta$ is the threshold noise level (e.g., 65 dBA, 70 dBA, …). $\delta$ is the indicator function equal to 1 if $L_{i,t} > \theta$, 0 otherwise.

### 3.4.2. Census tract-level Gini coefficient

In addition to estimating the total number of people exposed to aircraft noise, it is equally important to understand how this exposure is distributed across space – specifically, between census tracts at each time of day. This is especially relevant in the context of noise mitigation policies such as runway alternation, which aim to distribute the burden of noise exposure more equitably among surrounding communities.

To quantify this spatial disparity, we introduce the census tract-level Gini coefficient, a measure of inequality in the distribution of noise exposure across census tracts at each hour. Let $\mu_t$ be the mean exposure across census tracts at time $t$, and let $D$ denote the number of census tracts. The Gini coefficient at hour $t$ is given by: $G_\theta^t = \frac{1}{2D^2\mu_t}\sum_{i=1}^{D}\sum_{j=1}^{D}|E_{i,t}^{(\theta)} - E_{j,t}^{(\theta)}|$. This formulation captures the spatial inequality in noise exposure at each hour and helps assess whether certain areas consistently bear a disproportionate burden. A Gini coefficient close to 0 indicates that exposure is evenly shared across all regions at that hour, while a coefficient approaching 1 implies that exposure is heavily concentrated in a few regions.

## 4 RESULTS AND DISCUSSION

### 4.1. Technical validation of population estimates and aircraft noise data

#### 4.1.1. Validation of population estimates

To validate the de facto population estimates, we compared them with traditional census-based estimates, examining their spatiotemporal distribution. Unlike conventional static census methods, our approach utilizes dynamic mobile phone location data to capture real-time fluctuations in population density. This allows us to account for temporal variations – such as higher densities in commercial districts during the day and increased densities in residential areas at night. For a robust evaluation, we selected two representative districts: Gangnam, a major commercial area with a dense built environment, and Yangcheon, a predominantly residential district (Figure 3). Our analysis shows that in Yangcheon, evening population densities align closely with census-derived estimates, which reflect the expected nighttime residency patterns. In contrast, Gangnam exhibits significantly higher daytime densities, consistent with its commercial nature (Figure 4). These findings further substantiate the reliability of estimates, highlighting the importance of incorporating fine-grained temporal population data in assessing and mitigating the community impacts of aircraft noise.

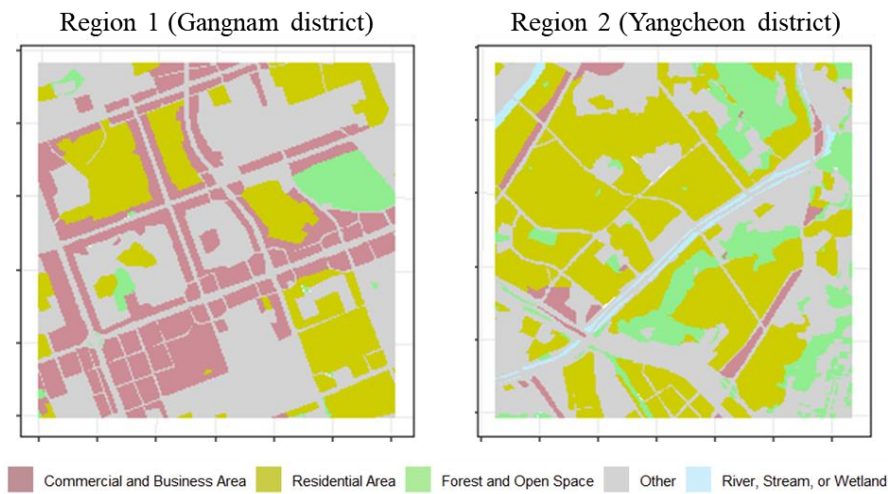

**Figure 3.** Land use distribution of Gangnam district (left) and Yangcheon (right)

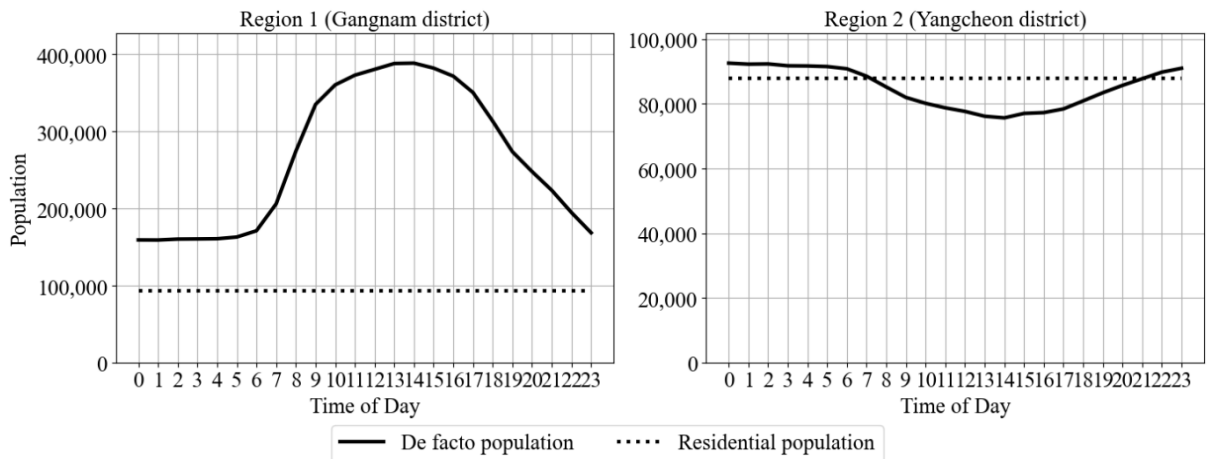

**Figure 4.** Hourly de facto and residential populations in Gangnam district (left) and Yangcheon district (right)

To further validate the de facto population estimates, we compared KT-provided data with a coarser dataset from SK Telecom (SKT), which offers publicly available hourly population counts at the district level. Since KT's data was originally available at the census tract level, we aggregated it to the district level to facilitate a direct comparison with SKT's subscriber-based counts. We compared the datasets by evaluating hourly absolute population counts and percentage changes from the previous hour. The results, shown in Figure 5, revealed strong positive correlations, with coefficients of determination of 0.99 for absolute counts and 0.95 for percentage changes. This high degree of agreement confirms that our approach accurately captures real-world population distributions.

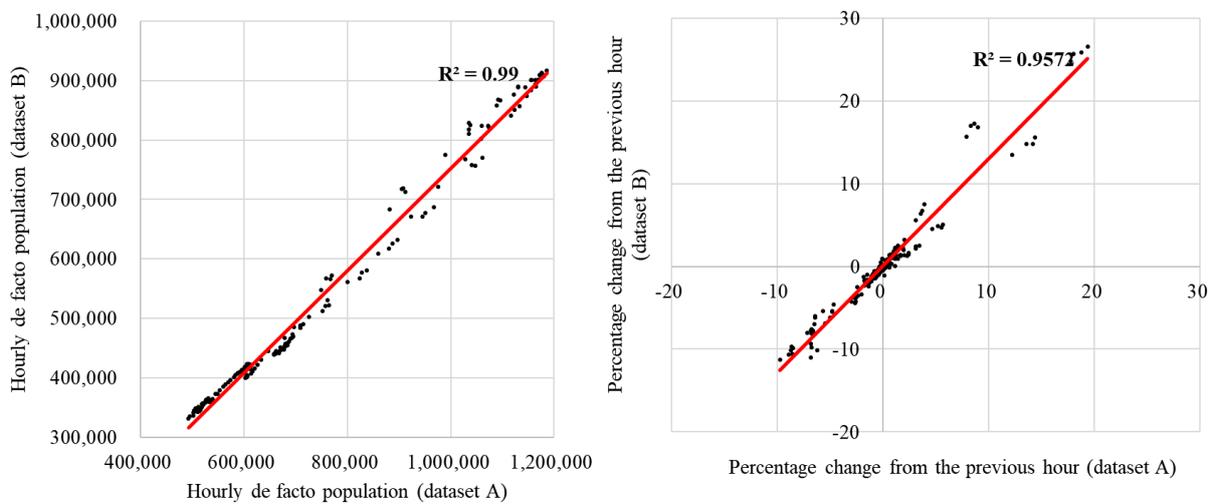

**Figure 5.** Comparison of hourly population counts and percentage changes between two datasets

### 4.1.2. Validation of airport noise estimates

We employed XGBoost to estimate noise levels near Gimpo Airport by integrating airport noise records, flight schedules, and meteorological data. The model achieved mean absolute errors of 1.74 for take-off noise and 2.01 for landing noise. Subsequent SHAP analysis revealed that key variables influencing noise levels are consistent with established domain knowledge, supporting the model's validity for explanatory analysis.

Figures 6 and 7 present the SHAP summary plots illustrating the impact of meteorological characteristics such as temperature, cloud cover, wind speed, and direction, on both take-off and landing noise levels. Among these factors, temperature emerged as the most significant in affecting both take-off and landing noise. Wind deviation showed a positive correlation with both take-off and landing noise, as indicated by higher wind deviation resulting in positive SHAP values (i.e. points towards the right are increasingly purple). Cloud cover also showed a positive correlation, consistent with related studies (Australian Government, 2020; TPA 2023, likely due to the noise reflection off clouds on cloudy days. Similarly, wind speed, particularly its direction, demonstrated a positive correlation with aircraft noise, aligning with its known effect on the acoustic footprint (Yunus et al., 2021).

Among Airbus aircraft and engine combinations, A330 with PW4, A320 with IAE V2500-A5, and A220 with PW1, exhibited varying degrees of impact on airport noise levels. Specifically, noise intensity increased with more frequent take-offs and landings involving A320 with IAE V2500 A5 engine. Moreover, Additionally, the number of A330 flights with PW4 engine correlated negatively with predicted noise levels, suggesting that a higher proportion of such aircraft in the fleet mix leads to lower noise levels; a higher number of such aircraft have negative SHAP values (i.e. points towards the left are increasingly purple). Similarly, Boeing aircraft, such as B737 with CFM 56, B767 with GE CF6-80, and B737 MAX with CFMI LEAP, had significant effects on both take-off and landing noise. Notably, take-off noise levels showed a positive correlation with higher frequencies of B737 with CFM56 and B767 with CFM6-80 flights. Additionally, the number of flights featuring the B737-8MAX with CFMI LEAP engine was found to correlate with landing noise levels. If such combination makes up a larger portion of flights occurring within an hour, the overall noise level near the airport is predicted to be lower.

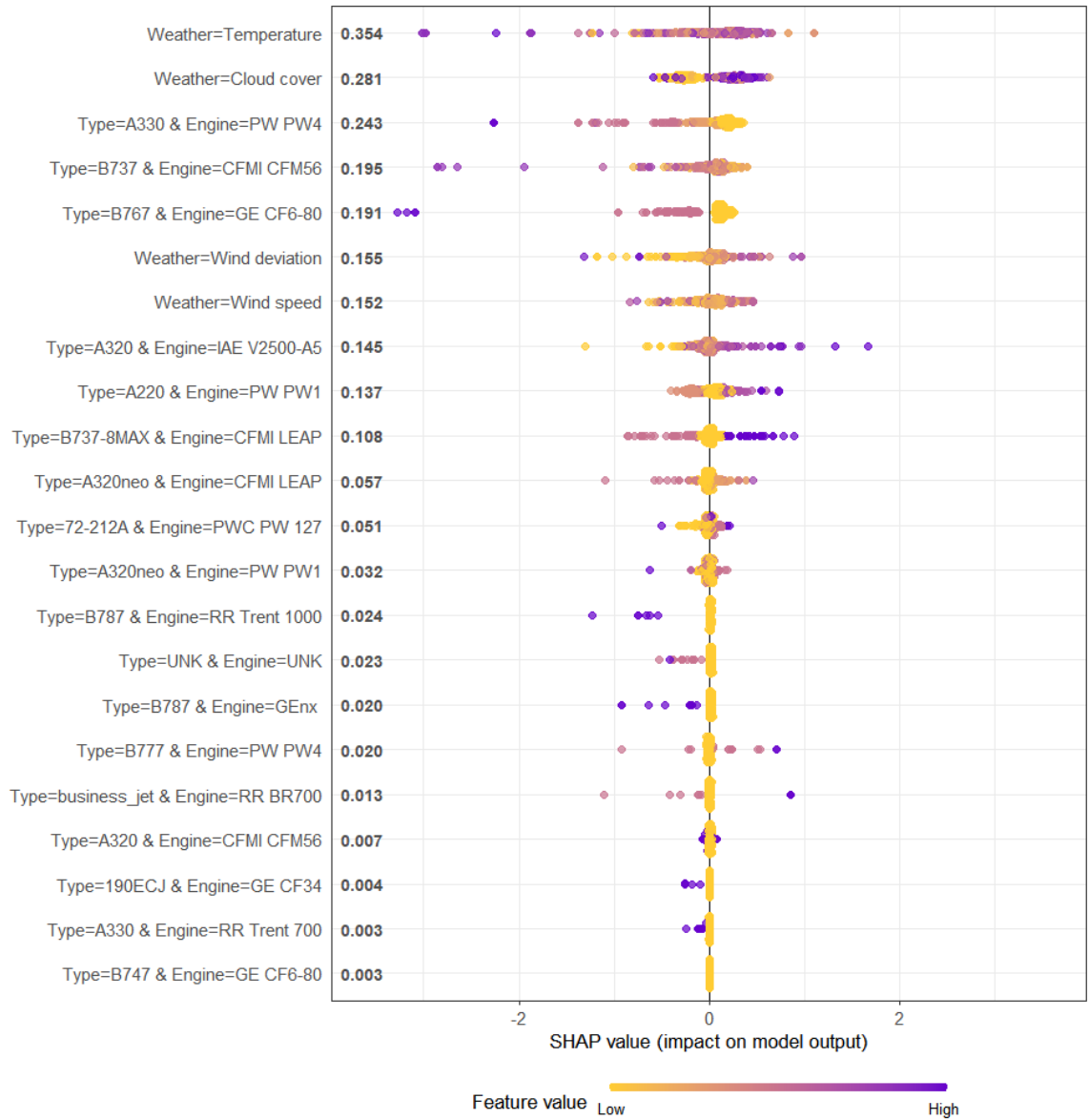

**Figure 6.** SHAP summary plot for take-off noise levels near the airport, ranked by mean absolute SHAP value

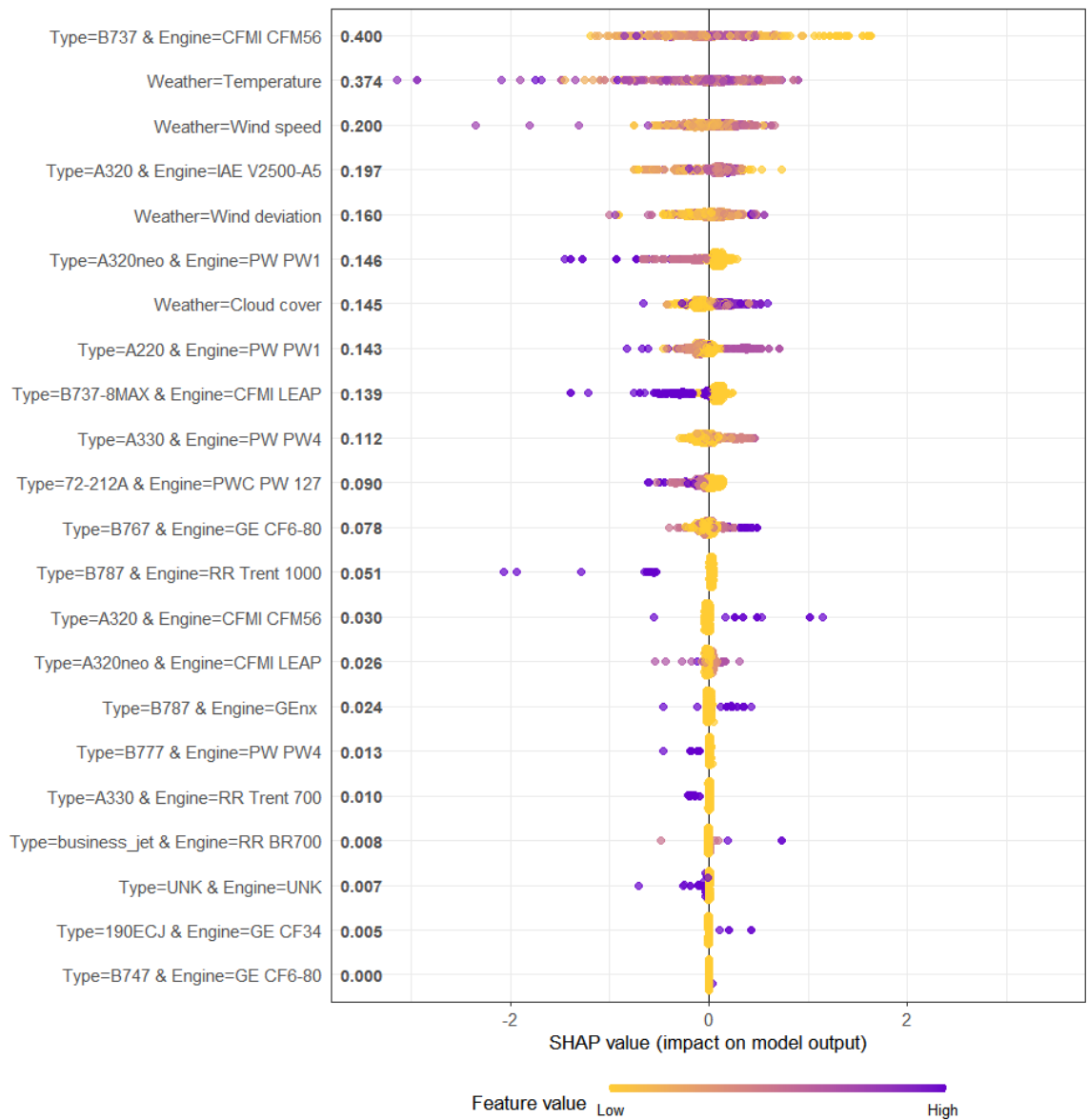

**Figure 7.** SHAP summary plot for landing noise levels near the airport, ranked by mean absolute SHAP value

We further explored the nonlinear relationships between dependent and independent variables using SHAP dependence plots. Figure 8 shows how meteorological factors such as cloud cover, wind speed, and wind deviation influence airport noise.

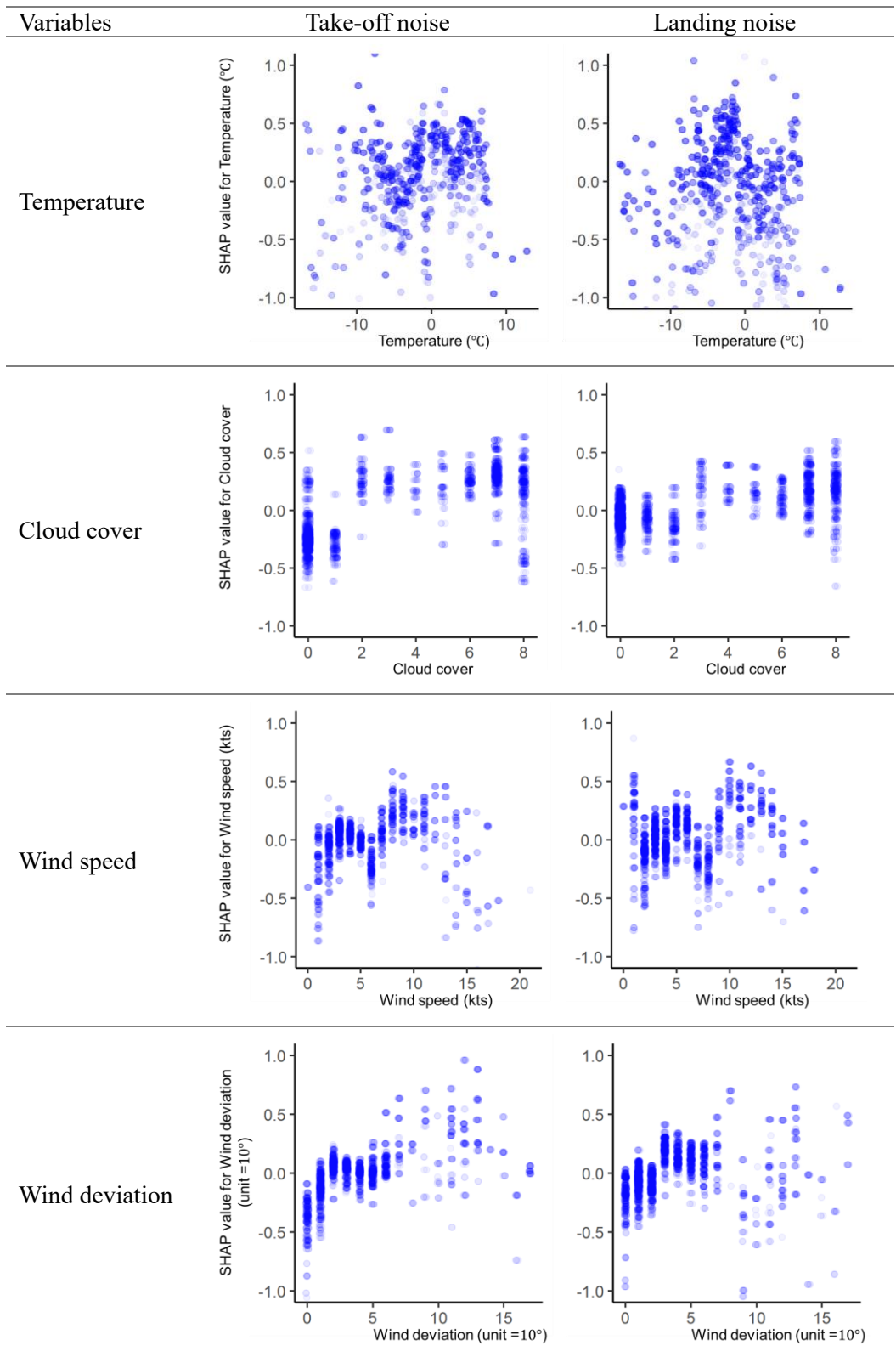

**Figure 8.** SHAP dependence plots of meteorological features

SHAP values above the $y = 0$ line indicate higher predicted noise levels, while those below suggest lower predictions. For instance, the variable values at which SHAP values turn positive were two cloud covers for take-off and three for landing, and seven knots of wind speed for take-off and nine knots for landing. Higher wind speeds typically resulted in increased aircraft noise; however, noise levels near the airport occasionally decreased due to sound dispersion caused by wind. This phenomenon could explain a slight drop in the observed noise trend depicted in Figure 8.

Meanwhile, a vertical dispersion in SHAP values is observed for some variables. This can be attributed to interaction effects with other variables, as SHAP value of an instance for a specific feature is not solely dependent on the value of that variable but is also influenced by the values of other variables associated with the instance. Though temperature is typically correlated with higher noise levels due to increased thrust generation by aircraft at higher temperatures (FAA, 2009), our data did not show significant trends, likely due to the cold winter season during data collection.

Figure 9 shows how the number of flights per hour, categorized by aircraft and engine combinations, affects hourly noise levels. Higher frequencies of B737 with CFM 56, A330 with PW4, and B767 with GE CF6-80 were associated with increased take-off noise. While engine specifications play a significant role, differences in climb rates among aircraft with the same engines can further influence the hourly noise level during take-off. For instance, GE CF6-80 engine, typically used on heavy aircraft like the B777 and B747, may allow for faster climb rates in relatively lighter aircraft such as the B767.

On the other hand, there is a clear correlation between an increase in flights involving A320 with IAE V2500-A5 engine and heightened take-off noise levels. Furthermore, nonlinear relationships are observed with A220 with PW1 and B737-8MAX with CFMI LEAP engine, where take-off noise is significantly impacted when the number of flights involving these aircraft is specifically two during their operating hours. Further analysis of the share of these aircraft in the total number of flights during specific time periods may provide additional insights into this relationship.

For landing noise, higher frequencies of flights involving A330 with PW4 and B767 with GE CF6-80 engine are associated with increased noise levels. Additionally, the relationship between the number of flights and landing noise is nonlinear for B737 with CFM56 and A320 with IAE V2500-A5 engine. Specifically, A320 with IAE V2500-A5 engine shows a notable increase in landing noise during certain time periods, while B737 with CFM56 engine tends to contribute to higher landing noise once flight frequency reaches a certain threshold. In contrast, flights involving B737-8MAX with CFMI LEAP and A320neo with PW1 engine show a negative correlation with landing noise. This can be attributed to the quieter nature of these newer aircraft, which are equipped with advanced engine technology. A higher proportion of flights featuring these aircraft types may contribute to lower predicted landing noise levels.

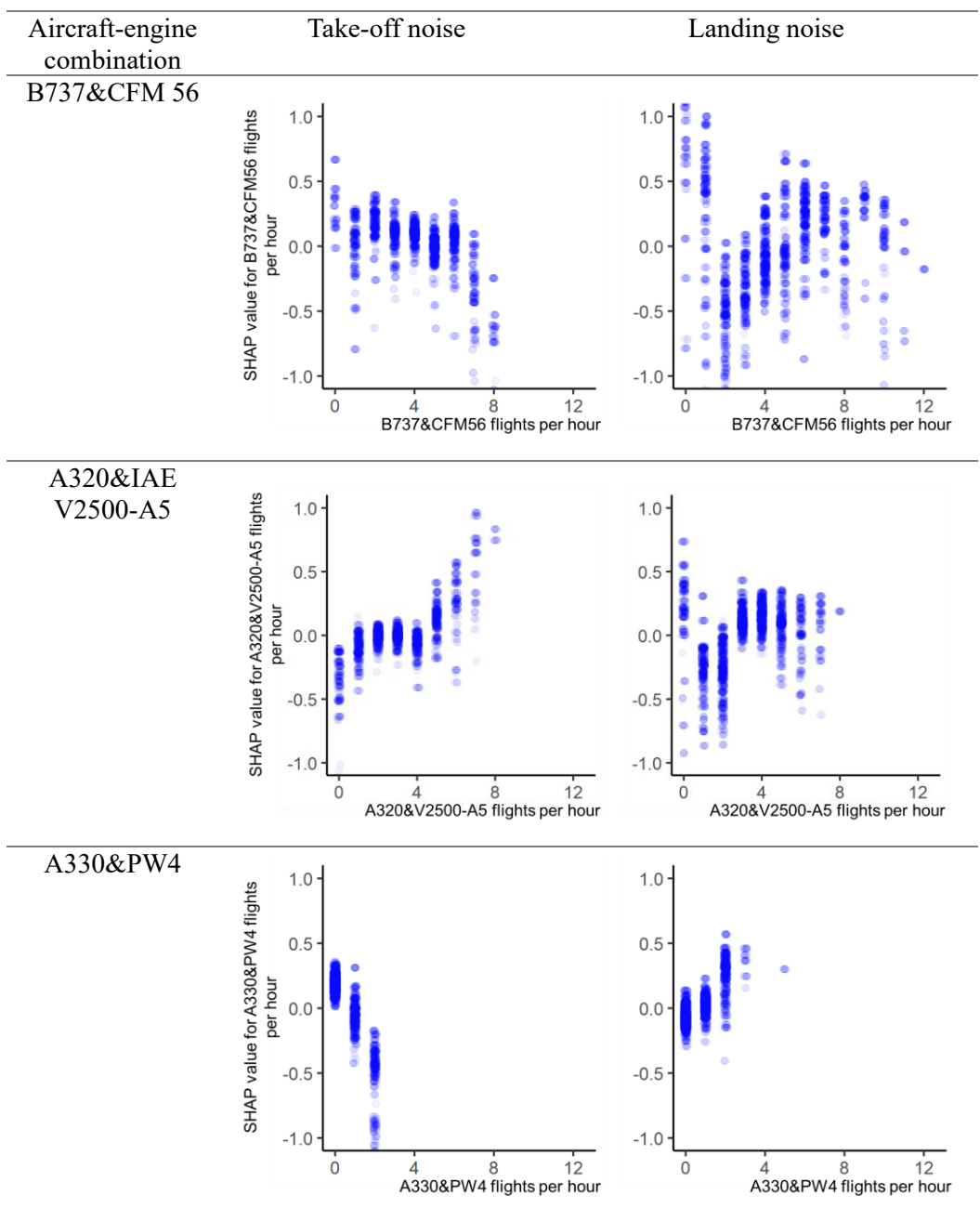

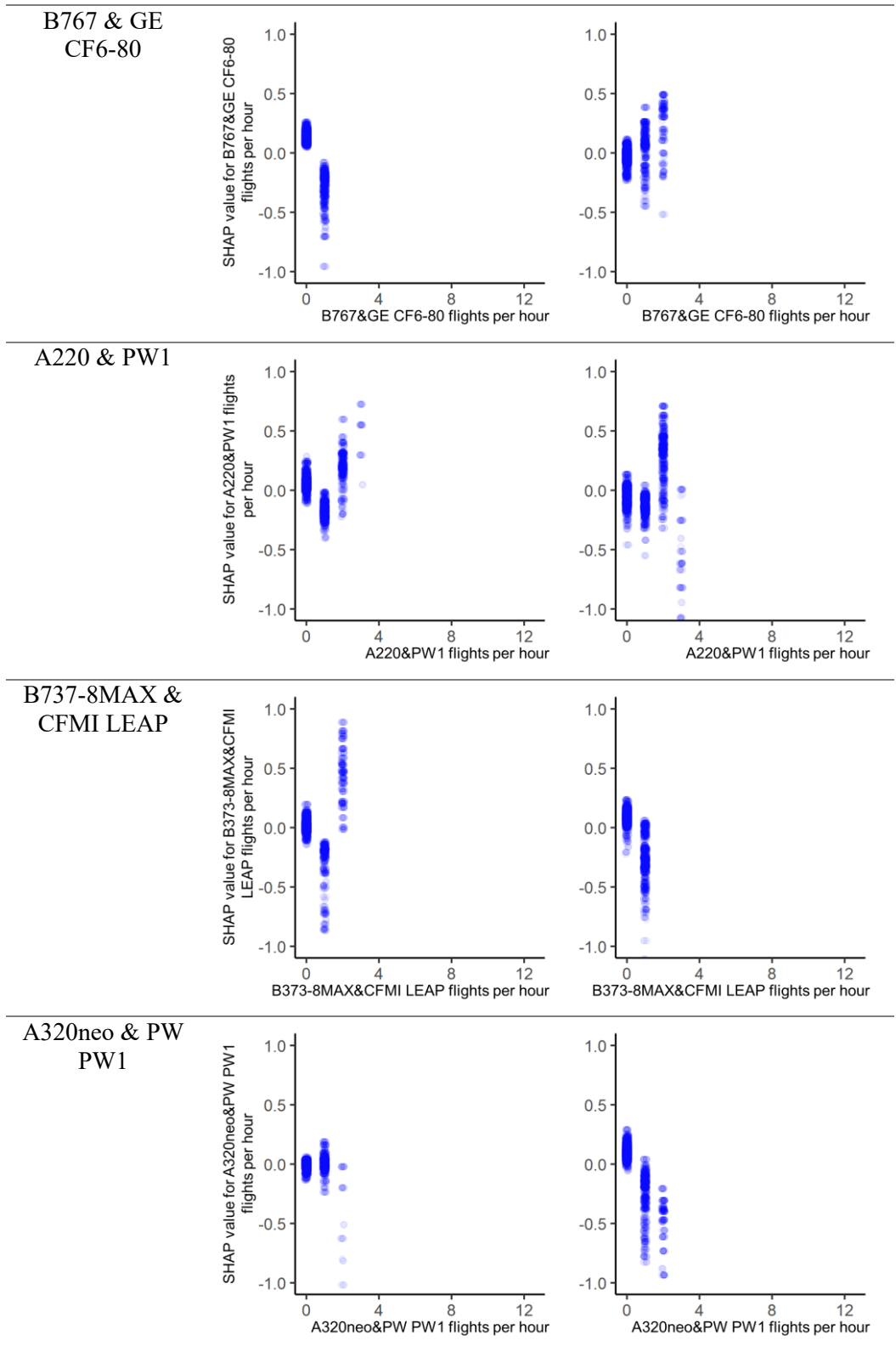

**Figure 9.** Selected SHAP dependence plots for hourly aircraft and engine fleet mix

## 4.2. Noise exposure analysis

Figure 10 presents a comparison between residential (static) and de facto (dynamic) population estimates and their respective noise exposure levels at two thresholds: 65 dBA and 70 dBA. The left plot shows total population counts over time for both residential (dotted black line) and de facto populations (solid black line). As expected, the residential population remains constant across all time slots, reflecting the static nature of census data. In contrast, the de facto population varies across time, peaking during morning and evening hours. This variation captures daily human mobility patterns, offering a more realistic representation of potential exposure during airport operations.

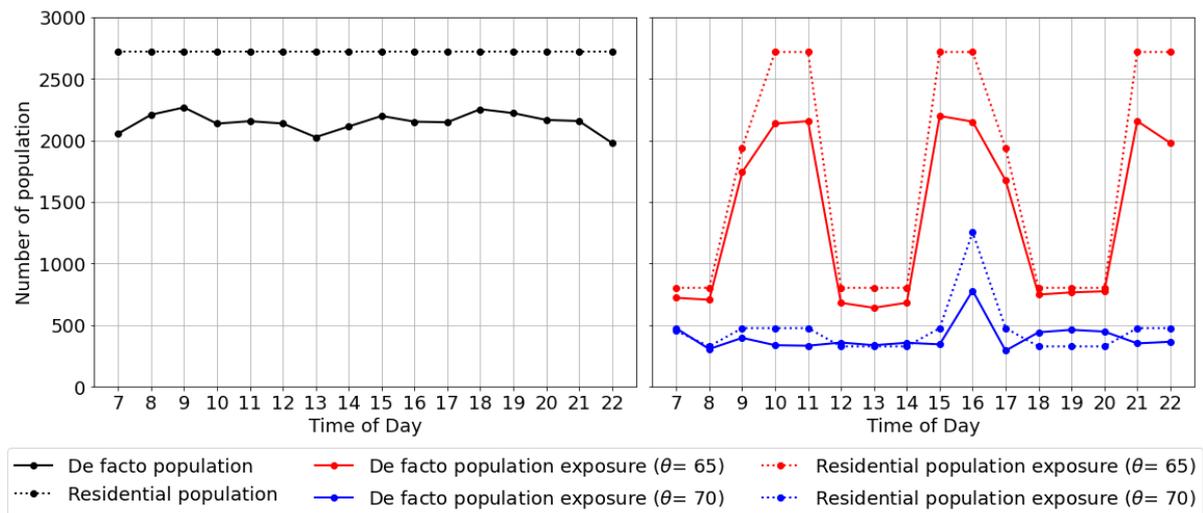

**Figure 10.** De facto and residential population in noise-affected areas (left) and exposure population at thresholds of 65 and 70 dBA (right)

The right plot shows the number of individuals exposed to aircraft noise for 65 and 70 dBA thresholds. In both cases, the number of people exposed according to de facto population data is consistently lower than when using residential population data. For instance, at 09:00, 1,741 individuals are exposed to noise above 65 dBA according to de facto population estimates, compared to 1,939 based on residential estimates. However, an important exception emerges for the 70 dBA threshold between 18:00 and 20:00, when the de facto-exposed population exceeds that of the residential baseline. This suggests that while many residents may have returned home by evening, high-noise regions experience transient population inflows (e.g., workers, commuters, or visitors) not captured in resident counts. Overall, these findings demonstrate that traditional census-based exposure metrics may misrepresent actual burden, either over- or under-estimating exposure depending on the hour and noise intensity. This underscores the value of integrating dynamic population data for more temporally and spatially accurate noise impact assessments.

Despite the relatively modest variation in total population throughout the day (a difference of 200-300 persons), the number of individuals exposed at higher noise thresholds shows substantial hourly fluctuation. For example, de facto exposure above 65 dBA ranges from 722 at 07:00 to 2,152 at 16:00. Similarly, exposure above 70 dBA ranges from 306 to 780. This suggests that variation in aircraft operations–such as frequency, aircraft type, and runway usage–plays a substantial role in shaping hourly exposure patterns, beyond what can be explained by population movement alone.

In addition, a subtle divergence is observed between the temporal patterns of 65 and 70 dBA exposure. Exposure above 65 dBA is more sensitive to operational variability, likely due to its broader spatial impact and greater overlap with mobile population zones. Many aircraft events exceed 65 dBA but not 70 dBA, and a larger portion of noise events cluster near the 65 dBA threshold. In contrast, exposure above 70 dBA is more spatially concentrated, confined to areas directly under flight paths.

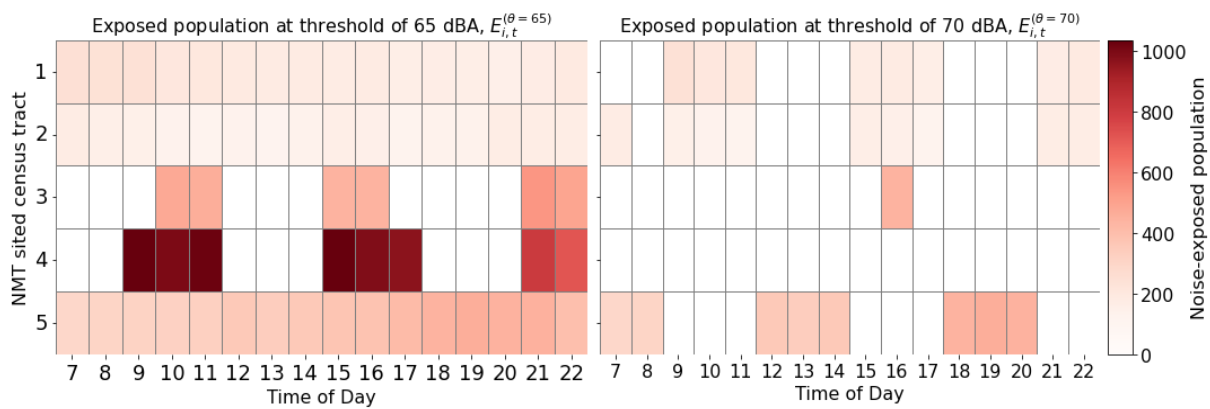

**Figure 11.** Hourly noise-exposed population by census tract at thresholds of 65 dBA (left) and 70 dBA (right)

Figure 11 presents heatmaps showing the hourly distribution of de facto population exposed to aircraft noise greater than or equal to 65 and 70 dBA across the selected census tracts. At the 65 dBA threshold, certain tracts exhibit sustained medium exposure throughout the day. Census tract 5, for example, experiences consistently moderate levels of exposure across almost all hours, reflecting its location under regular flight paths. Census tract 4 demonstrates a distinctive pattern, with very high exposure only from 09:00 to 12:00, from 15:00 to 18:00, and from 21:00 to 23:00, suggesting that specific operational activities (e.g., departures or landings) are affecting a narrowly defined area. Census tract 3 similarly shows pronounced exposure during specific periods.

At the 70 dBA threshold, exposure is more spatially and temporally restricted. Only a subset of the tracts shows non-zero exposure at this higher threshold. Census tract 5 experiences intermittent high exposure, while other tracts such as census tract 1 and 2 show scattered occurrences of high exposure, usually during specific hours. Several tracts that had exposure at 65 dBA, such as census tract 3 and 4, show no exposure above 70 dBA, indicating that they are located near the outer boundary of the 70 dBA contour and are only intermittently affected by high-intensity noise events.

These results confirm that high-level noise exposure is both spatially concentrated and temporally variable. The differences between 65 dBA and 70 dBA threshold patterns suggest that moderate noise levels affect a broader range of areas and populations, while higher noise levels are more localized and confined to specific zones during certain hours. This underscores the inadequacy of static exposure metrics and emphasizes the need for temporally resolved, spatially disaggregated exposure assessments.

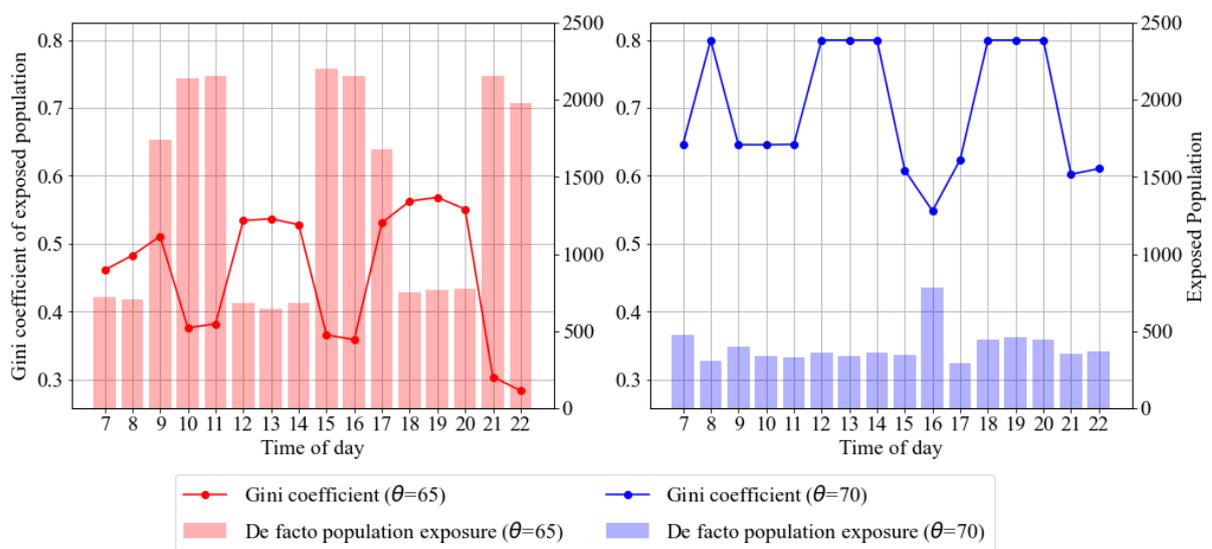

**Figure 12.** Gini coefficient of noise-exposed population by time of day at thresholds of 65 dBA (left) and 70 dBA (right)

To assess the spatial equity of aircraft noise exposure, we calculated the Gini coefficient of aircraft noise exposure across census tracts at each hour of the day, alongside the corresponding de facto population exposed to noise above 65 and 70 dBA (see Figure 12). The Gini coefficient measures the inequality of exposure distribution across census tracts, where higher values indicate greater concentration of noise burden in fewer areas.

For the threshold of 65 dB(A), the Gini coefficient fluctuates between 0.28 and 0.56, with a notable dip at 21:00–22:00, suggesting that exposure is more spatially even during the late evening. Conversely, higher inequality (e.g., 0.53–0.56) is observed in the late afternoon and early evening (17:00–19:00), even when the number of exposed individuals is moderate. This implies that during these hours, noise exposure is disproportionately concentrated in a smaller subset of tracts. At the 70 dB(A) threshold, spatial inequality is consistently higher, with Gini coefficients clustering around 0.60–0.80 throughout most of the operating hours. In particular, the coefficients reach 0.80 at 08:00, 12:00, 13:00, and 20:00, despite relatively low numbers of exposed population. This indicates that when high-level noise occurs, it is even more spatially concentrated—i.e., a few tracts bear the brunt of high-intensity noise, while others remain unaffected.

These results underscore the importance of not only tracking how many people are exposed, but also where the exposure is concentrated. For instance, similar total exposure levels (e.g., 359 at 12:00 vs. 344 at 15:00 for $\theta=70$) can lead to very different Gini coefficients (0.80 vs. 0.61), highlighting the uneven distribution of noise burden across space. Such temporal-spatial insights support policy evaluation of runway rotation schedules, noise abatement flight procedures, or zoning measures, especially when trying to ensure fair distribution of environmental burdens. Integrating both population magnitude and spatial inequality enables a more nuanced understanding of the lived noise experience and the equity dimension of aircraft noise exposure.

**4 CONCLUSIONS**

This study enhances current approaches to airport noise exposure assessment by integrating fine-grained noise monitoring with temporally resolved, mobile phone–based de facto population estimates. Unlike traditional methods that rely on static residential data, our approach captures intra-day variation in exposure by human mobility. The results show that static models tend to overestimate exposure during daytime hours when residents are absent from high-noise areas, and may also fail to detect peak burden periods when transient populations are present. By accounting for both spatial and temporal variation in population distribution, this method offers a more realistic and policy-relevant view of who is actually exposed—and when—enabling more equitable and responsive noise mitigation strategies.

The validation of our de facto population estimates, through comparison with traditional census-based data and another telecommunication provider's dataset, reveals a strong correlation, further confirming the reliability of our approach. Additionally, the use of machine learning models, specifically XGBoost, along with SHAP analysis, allows for interpretable predictions of airport noise, providing insights into the factors influencing noise levels. These insights are essential for understanding the complex interactions between aircraft operations, meteorological conditions, and noise propagation.

Our analysis of noise exposure patterns at different thresholds (65 dBA and 70 dBA) highlights temporal fluctuations in exposure, which is due to the dominant role of operational variability in shaping exposure patterns. Notably, while moderate noise levels (above 65 dBA) affect a broader set of areas, higher thresholds (above 70 dBA) are spatially concentrated in specific zones directly under flight paths. Furthermore, by calculating spatial inequity using the Gini coefficient, we reveal that noise exposure tends to be more concentrated in specific areas, especially at higher thresholds. This suggests that current noise management policies, which rely on aggregate exposure metrics, may miss critical disparities in exposure across different neighborhoods.

Our study area exhibited spatially rotating exposure patterns—that is, certain regions experience 3-hour blocks of higher exposure followed by 3-hour blocks of relative relief, with this alternating pattern mostly persisting over the course of a month (see Supplementary Figure S1). This operational structure constitutes a form of predictable respite, a concept identified as critical for community acceptance in the Heathrow study (Porter, 2023). A key contribution of this study is identifying which populations benefit from or are burdened by scheduled noise intervals. Our study addresses this gap by incorporating de facto population exposure modeling, which could be used to refine the concept of respite beyond the physical routing of aircraft. A time block may constitute respite operationally, but if it coincides with a peak in population presence, the net societal impact may be underestimated or misallocated. While the Heathrow studies demonstrated that predictable respite improves community tolerance, our findings indicate that population-adjusted scheduling may be critical to achieving both perceived and actual fairness. Absent such consideration, well-intentioned operational rotations may inadvertently concentrate noise burdens on more densely populated areas, undermining the policy's intended equity.

While this study advances noise exposure assessment through the integration of high-resolution mobility and acoustic data, several limitations should be acknowledged. First, the de facto population estimates rely on mobile phone signal data, which, despite adjustments, may underrepresent specific demographic groups such as children, the elderly, or individuals without mobile devices. Second, exposure was modeled based on data from a single month (January 2023), which may not reflect seasonal or operational variability in noise patterns. Future research should incorporate multi-month or annual datasets to capture more representative long-term trends. Third, while we used fixed monitoring terminal locations to approximate tract-level exposure, spatial interpolation or acoustic propagation modeling may be needed to fully capture local micro-scale variability. Lastly, this study focused on noise levels above fixed thresholds; incorporating dose–response health metrics or subjective annoyance responses could further enhance the policy relevance of the exposure indicators.

**CRediT authorship contribution statement**

**Soohwan Oh:** Conceptualization, Data curation, Formal analysis, Investigation, Methodology, Software, Validation, Visualization, Writing – original draft. **Hyunsoo Cho:** Data curation, Formal analysis, Investigation. **Jungwoo Cho:** Conceptualization, Formal analysis, Investigation, Methodology, Validation, Visualization, Writing – original draft.

**REFERENCES**

Ang, L. Y. L., & Cui, F. (2022). Remote work: Aircraft noise implications, prediction, and management in the built environment. Applied Acoustics, 198, 108978.

Australian Government. (2020). Understanding aircraft noise. Available online: https://www.westernsydneyairport.gov.au/sites/default/files/documents/2020-factsheet-understanding-aircraft-noise.pdf (Accessed on 29 June 2023).

Babisch, W., Houthuijs, D., Pershagen, G., Cadum, E., Katsouyanni, K., Velonakis, M., ... & HYENA Consortium. (2009). Annoyance due to aircraft noise has increased over the years – Results of the HYENA study. Environment international, 35(8), 1169-1176.

Chen, T., & Guestrin, C. (2016, August). Xgboost: A scalable tree boosting system. In Proceedings of the 22nd acm sigkdd international conference on knowledge discovery and data mining (pp. 785-794).

Ganic, E., Márki, F., & Schreckenberg, D. (2021). The population's daily movement and activities: Does it matter for aircraft noise impact assessment? *International Airport Review*, 25(01), 30-32.

Eagan, M. E. (2007). Supplemental metrics to communicate aircraft noise effects. Transportation research record, 2011(1), 175-183.

European Commission. (2020). Directive (EU) 2020/367 of the European Parliament and of the Council of 17 February 2020 amending Annex III of Directive 2002/49/EC relating to the assessment and management of environmental noise.

European Union. (2002). Directive 2002/49/EC of the European Parliament and of the Council of 25

June 2002 relating to the assessment and management of environmental noise.

Federal Aviation Administration. (2009). Pilot′s handbook of aeronautical knowledge. Skyhorse Publishing Inc.

Filippone, A. (2014). Aircraft noise prediction. Progress in Aerospace Sciences, 68, 27-63.

Franz, S., Rottoli, M. & Bertram, C. (2022). The wide range of possible aviation demand futures after the COVID-19 pandemic. Environmental Research Letters, 17(6), 064009.

Gagliardi, P., Teti, L., & Licitra, G. (2018). A statistical evaluation on flight operational characteristics affecting aircraft noise during take-off. Applied acoustics, 134, 8-15.

Ganic, E., Raje, F., & van Oosten, N. (2023) New perspectives on spatial and temporal aspects of aircraft noise: Dynamic noise maps for Heathrow airport. Journal of Transport Geography, 106, 103527.

Girvin, R. (2009). Aircraft noise-abatement and mitigation strategies. Journal of air transport management, 15(1), 14-22.

Grampella, M., Martini, G., Scotti, D., & Zambon, G. (2016). The factors affecting pollution and noise environmental costs of the current aircraft fleet: An econometric analysis. Transportation Research Part A: Policy and Practice, 92, 310-325.

Guo, Y., Thomas, R. H., Clark, I. A., & June, J. C. (2019). Far-term noise reduction roadmap for the midfuselage nacelle subsonic transport. Journal of Aircraft, 56(5), 1893-1906.

ICAO. (2008). Doc 9829: Guidance on the Balanced Approach to Aircraft Noise Management. International Civil Aviation Organization.

Jin, F., Li, Y., Sun, S., & Li, H. (2020). Forecasting air passenger demand with a new hybrid ensemble approach. Journal of Air Transport Management, 83, 101744.

Lefevre, M., Chaumond, A., Champelovier, P., Allemand, L. Gl., Lambert, J., Laumon, B., &Evrard, A. S. (2020). Understanding the relationship between air traffic noise exposure and annoyance in populations living near airports in France. Environment international, 144, 106058.

Lundberg, S. M., & Lee, S. I. (2017). A unified approach to interpreting model predictions. Advances in neural information processing systems, 30.

Ministry of Science and ICT. (2023). Results of the 2023 Telecommunication Service Coverage Inspection and Quality Evaluation. Available online: https://www.msit.go.kr/bbs/view.do?sCode=user&mId=113&mPid=238&bbsSeqNo=94&nttSeqNo=3183897 (Accessed on 4 April 2025).

Nguyen, D. D., Levy, J. I., Kim, C., Lane, K. J., Simon, M. C., Hart, J. E., ... & Peters, J. L. (2023). Characterizing temporal trends in populations exposed to aircraft noise around US airports: 1995–2015. Journal of Exposure Science & Environmental Epidemiology, 1-10.

Perron, S., Tetreault, L. F., King, N., Plante, C., &Smargiassi, A. (2012). Review of the effect of aircraft noise on sleep disturbance in adults. Noise and health, 14(57), 58.

Porter, N. (2017). Respite from aircraft noise: Overview of recent research. Heathrow Airport Limited.

Porter, N. (2023). Respite from aircraft noise: Summary of research journey. Heathrow Airport Limited.

Postorino, M. N., & Mantecchini, L. (2016). A systematic approach to assess the effectiveness of airport noise mitigation strategies. Journal of Air Transport Management, 50, 71-82.

Revoredo, T., Mora-Camino, F., & Slama, J. (2016). A two-step approach for the prediction of dynamic aircraft noise impact. Aerospace Science and Technology, 59, 122-131.

Seoul Metropolitan Government (2025). Seoul living population. Available online: https://data.seoul.go.kr/dataVisual/seoul/seoulLivingPopulation.do (Accessed on 4 April 2025).

Simons, D. G., Besnea, I., Mohammadloo, T. H., Melkert, J. A., & Snellen, M. (2022). Comparative assessment of measured and modelled aircraft noise around Amsterdam Airport Schiphol. Transportation Research Part D: Transport and Environment, 105, 103216.

Tenney, A. S., Glauser, M. N., Ruscher, C. J., & Berger, Z. P. (2020). Application of artificial neural networks to stochastic estimation and jet noise modeling. AIAA journal, 58(2), 647-658.

Tirtha, S. D., Bhowmik, T., & Eluru, N. (2022). An airport level framework for examining the impact of COVID-19 on airline demand. Transportation Research Part A: Policy and Practice, 159, 169-181.

Tirtha, S. D., Bhowmik, T., & Eluru, N. (2023). Understanding the factors affecting airport level demand (arrivals and departures) using a novel modeling approach. Journal of Air Transport Management, 106, 102320.

Tveter, E. (2017). The effect of airports on regional development: Evidence from the construction of regional airports in Norway. Research in Transportation Economics, 63, 50-58.

Vela, A. E., & Oleyaei-Motlagh, Y. (2020, October). Ground level aviation noise prediction: a sequence to sequence modeling approach using LSTM recurrent neural networks. In 2020 AIAA/IEEE 39th Digital Avionics Systems Conference (DASC) (pp. 1-8). IEEE.

World Health Organization. (2018). Environmental Noise Guidelines for the European Region. World Health Organization Regional Office for Europe.

Yunus, F., Casalino, D., Avallone, F., & Ragni, D. (2021). Toward inclusion of atmospheric effects in the aircraft community noise predictions.

Zellmann, C., Schäffer, B., Wunderli, J. M., Isermann, U., & Paschereit, C. O. (2018). Aircraft noise emission model accounting for aircraft flight parameters. Journal of Aircraft, 55(2), 682-695.